# Excitonic lasing of strain-free InP(As) quantum dots in AlInAs microdisk

D. V. Lebedev[1], M. M. Kulagina[1], S. I. Troshkov[1], A. A. Bogdanov[2], A. S. Vlasov[1], V. Y. Davydov[1], A. N. Smirnov[1], J. L. Merz[2], J. Kapaldo[3], A. Gocalinska[4], G. Juska[4], S. T. Moroni[4], E. Pelucchi[4], D. Barettin[5], S. Rouvimov[1,2] and A. M. Mintairov[1,2]

[1] *Ioffe Physico-Technical Institute, Russian Academy of Sciences, Saint Petersburg 194021, Russia*
[1] *ITMO University, Saint Petersburg 199034, Russia*
[2] *Electrical Engineering, University of Notre Dame, Notre Dame, Indiana 46556, USA*
[3] *Physics Department, University of Notre Dame, Notre Dame, Indiana 46556, USA*
[4] *Tyndall National Institute, University College Cork, Ireland*
[5] *"Università Niccolo Cusano" 00133 and University of Rome "Tor Vergata" 00166, Italy, Rome*

Formation, emission and lasing properties of strain-free InP(As)/AlInAs quantum dots (QDs) embedded in AlInAs microdisk (MD) cavity were investigated using transmission electron microscopy and photoluminescence (PL) techniques. In MD structures, the QDs having nano-pan-cake shape have height of ~2 nm, lateral size of 20-50 nm and density of ~$5 \times 10^9$ cm$^{-2}$. Their emission observed at ~940 nm revealed strong temperature quenching, which points to exciton decomposition. It also showed unexpected type-I character indicating In-As intermixing, as confirmed by band structure calculations. We observed lasing of InP(As) QD excitons into whispering gallery modes in MD having dimeter ~3.2 $\mu$m and providing free spectral range of ~27 nm and quality factors up to Q~13000. Threshold of ~50 W/cm$^2$ and spontaneous emission coupling coefficient of ~0.2 were measured for this MD-QD system.

The demonstration of semiconductor microdisk (MD) cavity laser design[1] and realization of quantum dot (QD) active media in MDs[2-4] have generated a versatile monolithic solid-state nano-photonic platform which has been intensively developed during a few decades. Thanks to the flexibility of using different material systems[5-12], research topics covered fundamental themes on cavity quantum electrodynamics[5,13,14], cavity spin dynamics[15] and quantum optics,[16] together with device applications including single photon sources,[17] creation and optimization of laser diodes,[18,19] micro-photonic circuits[20,21] and mode control.[22]

The MD-QD platform jointly exploits the properties of cavity whispering gallery modes (WGM) propagating along the disk circumference, and the discrete, atomic-like energy spectrum of QDs which arises from electron and hole quantum confinement. This system offers very high cavity quality factors (e.g. up to $10^5$ in Ref. 12). It also allows for easy fabrication and photonic circuit integration procedures, with a strong localization of the excitonic and electron-hole recombination transitions, which provides screening from non-radiative and interface defects, together with high temperature operation and spectral stability. High optical gain of epitaxial QDs delivered nanoWatt laser threshold[23] and single dot lasing.[24] All in a wide spectral range from ultra-violet (310-360 nm) for GaN/AlN[10,11,25] QDs to infra-red (1400 and 1600 nm) for InAs/GaAs[14] and Ge/Si[7] QDs. Other material systems emitting inside this range are self-organized InGaN/GaN,[12] CdSe/ZnSe,[6] InP/GaInP[26,5] and GaAs/GaSb[9] QDs, and lattice-matched GaAs/AlGaAs QDs resulting from fluctuations in the quantum well thickness.

Recently unusual nanostructures of "*lattice matched*" InP on AlInAs were reported.[27] They are characterized as flat QD having a height of 8 nm, lateral size of 20-250 nm and density of ~$10^8$ cm$^{-2}$. They were observed by atomic force microscopy after deposition of a few nm thick InP on an AlInAs layer using metal organic vapor-phase epitaxy (MOVPE). These QDs have a combination of interesting features, which are new for the MD-QD system and can be exploited for device application and fundamental research. First, they are expected to have a type-II band alignment,[27,29] previously studied, for example, in GaAs/GaSb[9] QDs. Second, they have a large size, which has the potential to provide extremely strong photon-matter interaction[5] and low laser thresholds [23]. Finally, they can provide an easy stacking due to the lattice-matched conditions, which gives an additional possibility for engineering the active media. Open questions for using these QDs in MD cavities, however, stay with their structural uniformity - etching experiments have shown the presence of As at QD center, with effects of the P-As inter-mixing during capping to form the waveguide and with fundamentally poorly characterized emission properties till now.

Here we report on structural and photoluminescence (PL) spectroscopy studies of these QDs embedded in 250 nm thick waveguide structures, and on the lasing properties of the MDs fabricated from these structures. We also used band structure calculation to estimate/identify the supposed In-As intermixing. We found that capping of these QDs leads to a decrease of their overall height down to 2 nm forming an array of QDs having lateral sizes of 20-50 nm and the density of ~$5 \times 10^9$ cm$^{-2}$. These QDs show very strong emission peak at ~940 nm at 10 K, they show decay time of ~1 ns strongly indicating type-I band alignment. This probably results from As intermixing. This is supported by our calculations which suggest an As content of ~25%.

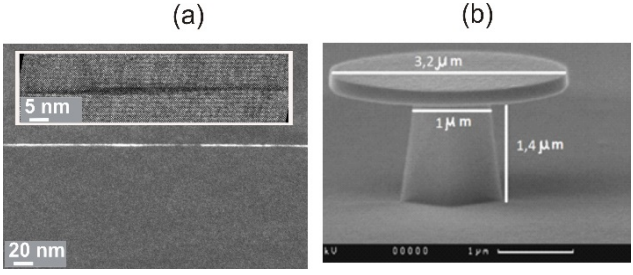

FIG. 1. Cross-section transmission microscopy images (the insert is of high resolution) of an InP(As)/AlInAs QD in a MD structure (a) and scanning electron microscopy image of AlInAs microdisk (b).

The QD emission is strongly quenched at higher temperature indicating excitonic emission. For the MDs we observed WGM lasing at temperatures <120 K having threshold of ~50 W/cm$^2$ at 10 K. We obtained quality factors of ~13 000 and spontaneous emission coupling coefficient $\beta$=0.23.

The details of the MOVPE growth process are described in Ref. 27. Here we used a semi-insulating InP substrate and grew a 100 nm thick InP buffer layer, 250 nm thick AlInAs waveguide having InP QDs in the middle. The AlInAs bottom layer was grown at a temperature of 600° C and the cap at 560° C (real estimated growth temperature) and with a growth rate of ~1 $\mu$m/h; thus the overgrowth of the InP QDs by a 125 nm thick AlInAs cap continued for ~7.5 minutes. V/III ratio was 110. We used undoped and n-type doped AlInAs layers. The n-type doping was achieved by ~3x10$^{17}$cm$^{-3}$ silicon doping of 20 nm part of AlInAs at distance of 20 nm from QDs. In preliminary studies we also used structures having a thin AlInAs cap of ~25 nm with the same overall structure.

For cross-section transmission electron microscopy (TEM) measurements (see Fig.1a) we used a FEI Titan 80-300 electron microscope. TEM samples were prepared by Focus Ion Beam method using FEI Helious Dual Beam SEM/FIB Nanofactory. The mushroom type MDs having diameters 2-4 $\mu$m (see scanning electron microscopy image in Fig.1b) were fabricated by optical photolithography and wet selective chemical etching using HCl, H$_3$PO$_4$: CH$_3$COOH.

Photoluminescence (PL) spectra were excited by a continuous-wave (CW) solid state ($\lambda$=532 nm) or by a pulsed ($\lambda$=635 nm, frequency 50MHz) laser and measured using a standard variable temperature $\mu$-PL set up having single-photon correlation capability.[30] The optical power on the sample was regulated by a set of neutral density filters.

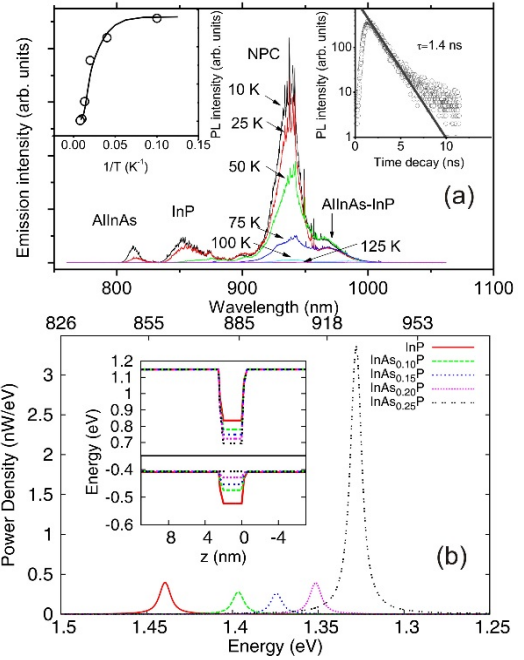

FIG. 2 Experimental (a) and calculated (b) emission spectra of InP(As)/AlInAs NPC-QD structures. Inserts in (a) are the intensity of a representative QD band peak versus temperature - left and time resolved photoluminescence– right; dots - experiment, curve – fit (see text). Spectra in (a) are for T=10, 25, 50, 75, 100 and 125 K; spectra in (b) are for As composition of x=0, 0.05, 0.1, 0.15, 0.2 and 0.25. Insert in (b) shows conduction and valence band potential along the vertical direction.

In the cross-section TEM images (see Fig.1a) the InP deposit in MD structure are seen as a white broken stripe having thickness of ~2 nm, which represent nano-pan-cake (NPC) shape QDs having density of ~5x10$^9$ cm$^{-2}$ and lateral size of ~50 nm (see the high resolution image in the insert). The thickness and the lateral size of capped dots decrease, while the density increases compared to uncapped samples.[27] From TEM data presented in Supplementary Materials (SM) we found that, in the structures having thin caps, the QDs were only slightly modified and had slightly reduced height (down to 5 nm from the ~8 nm in uncapped samples) and increased size (up to 300 nm). This indicates some form of "melting" of these NPC QDs and a planarization of the InP deposit induced by capping, at least with the here discussed growth conditions, with non-continuous areas of a "wetting layer" like structure too.[31]

In low-temperature (10K) PL spectra of unprocessed planar MD structures we observed emission bands from the NPC-QD ensemble at 940 nm, from the InP substrate at ~875 nm, from the InP-AlInAs interface at ~1000 nm[28,29] and from AlInAs at 810 nm. We also observed a feature around 860 nm, which could be possibly linked to unaffected (non-arsenised) InP nanostructures, even if it is not fully clear at this stage. The QD band can be hardly distinguished from the InP-AlInAs interface band, as the latter is broad and has order of magnitude stronger intensity. On the other hand, the interface band is strongly inhibited in MDs due to the mushroom shape and processing and the QD band in MDs can be clearly resolved. Noticeably, when the pumping power density is increased, the InP-AlInAs interface band shows a strong blue shift as expected for a type-II band alignment, while the QD band does not, indicating a possible type-I band alignment. The type-I band alignment is also consistent with the 1.4 ns emission decay lifetime[32] of QDs (see the left insert in Fig.2a), measured at 940 nm. In Fig.2a, we plotted the PL spectra of a 4 μm diameter MD at T=10, 25, 50, 75, 100 and 125 K under CW excitation. It is seen that when the temperature is increased, the QD band is strongly quenched – its intensity drops by two orders of magnitude in the temperature range 10-100 K and it disappears in the PL spectra at T>120 K. The measured intensity versus temperature is plotted in the insert in Fig.2a on a logarithmic scale versus reciprocal temperature (Arrhenius plot) and fitted by an expression describing the thermal activation of $i$ non-radiative channels having activation energies $E_i$ and capture times $\tau_i$[33]

$$I_{PL}(T) = I_0\left[1 + \sum_i \frac{\tau_0}{\tau_i} e^{-E_i/kT}\right]^{-1}, \quad (1)$$

where $\tau_0$ =1.4 ns is the radiative life-time. The fit gives two quenching states having $E_i$=7 and 40 meV and $\tau_i$=280 and 0.5 ps, which are responsible for the PL intensity decrease in the range below and above 100 K, respectively. The obtained $E_i$ values correspond well to exciton binding energies in QDs[34] and heavy-light hole splitting[35]: we can then infer that the strong temperature quenching of the emission of QDs is related to an exciton decomposition process.

The observed type-I exciton emission of QDs is an indication of P-As intermixing, which we modelled using band structure calculations in the framework of a three-dimensional 8-band k·p model[36]. An 8 × 8 effective-mass Hamiltonian has been implemented according to Foreman's application of Burt's exact envelope function theory to planar heterostructures.[37,38] The electromechanical field (Strain and Piezoelectric fields) have been calculated by a fully-coupled continuum model, described in details in Ref. 39. Piezoelectric and strain fields have been included in k·p via deformation potentials[40] with parameters derived from Refs. 41, 42.

Optical transitions have been calculated by Fermi's Golden Rule where the coupling matrices have been computed in the dipole approximation, evaluating the dipole matrix elements. The matrix elements are computed in the Heisenberg representation from the momentum matrix elements.[43] All simulations have been implemented and solved using the TiberCAD simulator.[44] In our calculations we simulated five different As content x in the QDs - x=0, 0.10, 0.15, 0.20 and 0.25. The average concentration within the dot has been considered in the context of a continuous model, without considering for any As local fluctuations.[45] Concerning the shape of the dot, we opted for a cylinder with a radius of 50 nm and height of 2 nm. The results of calculations of the optical transition strength together with conduction and valence band vertical potential distributions presented in Fig.2b reveal transition from a type-II to type-I band alignment at x=0.25, which results in an order of magnitude increase of emission intensity. The emission wavelength gradually increases from 860 to 935 nm when x changes from 0 to 0.25 reaching the experimentally observed values. Interestingly for the structures having thin cap the QD emission band is observed at 810-910 nm (see SM), which indicates x~0.1, i.e. smaller intermixing. Due to a nearly flat valance band the hole is nearly free at x=0.25 (see SM) and the type-I band alignment is indeed enhanced by the Coulomb interaction, i.e. the formation of an exciton (see Ref. 32).

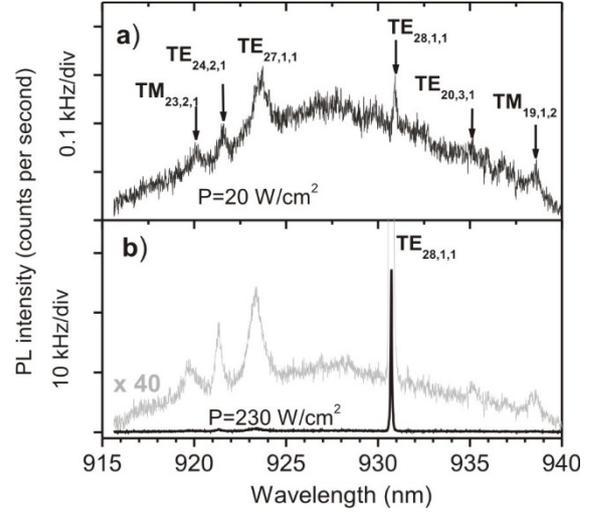

FIG. 3 μ-PL spectra taken under pulse excitation of a representative InP(As)/AlInAs microdisk (diameter 3.2 μm) for power density 20 and 230 W/cm² for (a) and (b), respectively.

The emission spectrum of QDs in MDs at low pumping power density reveals few sharp lines and peaks, related to a WGM, superimposed on the background QD ensemble band. For MD presented in Fig 3a the emission spectrum taken at 20 W/cm² average pulse excitation power contains at least five WGMs observed at 920.0, 921.6, 923.6, 931 and 938.7 nm and the background band has nearly two time larger peak intensity than WGMs. A full width at half maximum (FWHM) of WGMs has value 0.2-1 nm. Our calculations using COMSOL Multiphysics program[46] have shown (see details in SM) that the sharpest line observed at 931 nm can be attributed to $TE_{28,1,1}$ and that free spectral range for this MD is 26.6 nm.

Using time resolved experiments we obtained (see SM) that the emission decay of the WGM lines (0.8 ns) is faster than that of the ensemble band (1 ns), which is an indication that it results from a modification of the photon density of states induced by the cavity, i.e. what is known as Purcell effect.[47] Thus we can estimate the spontaneous emission fraction that is coupled into the mode, which, according to ref. 48, is $\beta=1-\tau_{cav}/\tau_{free}$, where $\tau_{cav}$ and $\tau_{free}$ are the emission decay times into the mode and into free space, respectively. This gives $\beta$ equal to 0.2.

At high pumping power density (see spectra for P=230 W/cm$^2$ in Fig.3b) we observed a two orders of magnitude increase of the emission intensity of TE$_{28,1,1}$ mode, from two to five times increase of other WGMs and no changes of emission intensity of background ensemble band, which manifests into clear lasing.

The dependence of the mode intensity on the input power density is plotted in Fig.4a in double logarithmic scale and shows a weak bending near 50 W/cm$^2$ indicating lasing effect. Such dependence is typical for microlasers having small modal volume.[49] For CW

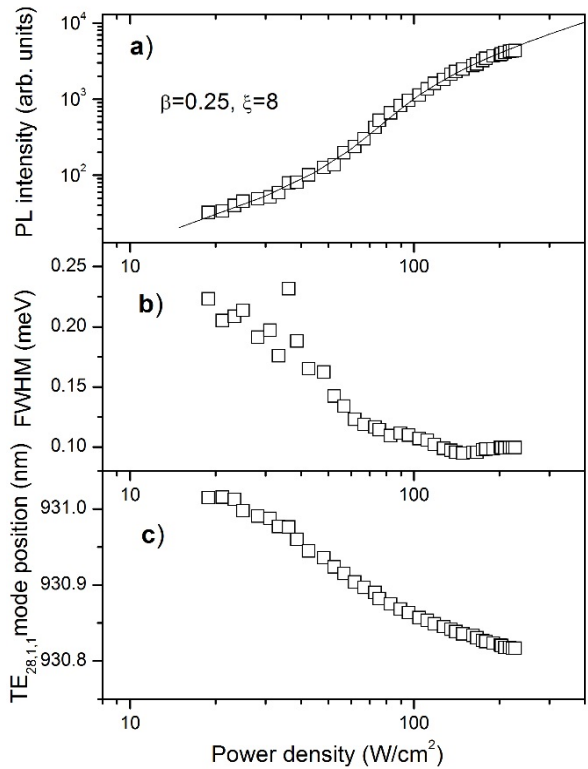

FIG.4 Power density dependence of TE$_{28,1,1}$ mode intensity (a), half-width – FWHM (b) and emission wavelength (position) (c).

excitation this dependence can be described by the expression:

$$I(p) = \frac{q\gamma}{\beta}\left[\frac{p}{p+1}(1+\xi)(1+\beta p) - \xi\beta p\right], \quad (2)$$

where $I$ is the excitation current and $p$ is the number of photons in the mode, $q$ is electron charge, $\gamma$ is the cavity decay rate, $\beta$ is e spontaneous emission coupling coefficient; $\xi$ is the ratio of spontaneous photon emission into the lasing mode and the cavity decay rate. It is defined as: $\xi = N_0\beta V/\gamma\tau_{sp}$, where $N_0$ is the transparency carrier concentration of the gain material, $V$ is the volume of active material, and $\tau_{sp}$ is the spontaneous lifetime of the active material. For optical excitation $I$ corresponds to the pumping power and $p$ is proportional to the output intensity. We used this expression to fit our data in Fig.4a accounting for the fact that the solution of rate equations of Ref. 49 for pulse excitation can be approximated by expression (2) assuming quasi-CW conditions during a single pump-emission event. This implies constant values of carrier density and photon number during the emission decay time under the condition of conservation of their total number. Varying $\beta$ near value of 0.2, obtained from time resolved measurements, we found $\beta=0.23$ and $\xi=8$ using expression (2) to fit the experimental data in Fig.4a.

The threshold power $P_{th}\sim 50$ W/cm$^2$ was estimated using an onset of a slope changing of the output-input curve.

The transition to lasing regime is accompanied by two time decrease of the FWHM of the mode line (see Fig.4b). The saturation point of FWHM decrease (~60 W/cm$^2$) seen in Fig.4b is in a good agreement with a $P_{th}$ value obtained from the input-output curve. The quality factor obtained above threshold for this MD is Q~10 000. The largest Q observed using different MDs is 13000.

In the SM are also presented measurements of the second-order correlation function g$^2$(0)=1.01±0.02 which, ideally, represents the expected coherent light emission, but this interpretation needs further investigations.

Increasing of the pump power results also in a blue shift of the mode position (Fig. 4c), which indicates a decrease of the index of refraction induced by the electron-hole/exciton gas.[50] We should point out that the measured $P_{th}$ and $\beta$ of our AlInAs MDs with InP(As) NPC-QDs are similar to that measured for GaInP MDs with self-organized InP QDs in Ref. 26.

In conclusion we investigated structural, emission and lasing properties of strain-free QD InP(As)/AlInAs nanostructures embedded in AlInAs MDs. We found that in MD structures these QD nanostructures have height ~2 nm, lateral size 20-50 nm and density ~5x10$^9$ cm$^{-2}$. The emission wavelength of these QDs at T=10 K is ~940 nm and has type-I character, which according to band structure calculations, corresponds to an As content of 0.25. The emission of QDs has strong temperature quenching due to exciton decomposition (activation energy of 7 meV). In AlInAs microdisk cavity having Q~10000 we observed lasing of InP(As) QDs into whispering gallery modes having threshold of ~50 W/cm$^2$ and spontaneous emission coupling coefficient of ~0.2.

A. M. M. and D.V.L. acknowledges support of the Ministry of Education and Science of the Russian Federation (contract № 14.Z50.31.0021, 7th April 2014). E.P., A.G, G.J, S.T.M acknowledge funding provided by Science Foundation Ireland under grants 12/RC/2276 and 10/IN.1/I3000. D.B acknowledges support of EC through grant No. 612600 LIMACONA and the Mediterranean Institute of Fundamental Physics.